\documentclass{article}
\usepackage{amssymb}
\usepackage{graphicx}
\usepackage{graphics}
\usepackage{epsfig}
\usepackage{psfrag}

\def\qed{\leavevmode\unskip\penalty9999 \hbox{}\nobreak\hfill
     \quad\hbox{\leavevmode  \hbox to.77778em{%
               \hfil\vrule   \vbox to.675em%
               {\hrule width.6em\vfil\hrule}\vrule\hfil}}
     \par\vskip3pt}

\newtheorem{proposition}{Proposition}
\newcommand{\proof}[1]{{\bf Proof: } #1 \qed}

\newcommand{\eq}[1]{\begin{equation}#1\end{equation}}
\newcommand{\tr}[1]{{\rm tr}\left[#1\right]}

\begin{document}
\title{\sc{Entanglement and Frustration in
Ordered Systems}}

\author{
 {M.M. Wolf, F. Verstraete, and J.I. Cirac}\\
  \footnotesize{Max-Planck-Institut f\"ur
Quantenoptik, Hans-Kopfermann-Str. 1}\\ \footnotesize{85748
Garching,  Germany}}
\date{{\small\today}}
\maketitle

\begin{abstract}
This article reviews and extends recent results concerning
entanglement and frustration in multipartite systems which have
some symmetry with respect to the ordering of the particles.
Starting point of the discussion are Bell inequalities: their
relation to frustration in classical systems and their
satisfaction for quantum states which have a symmetric extension.
It is then discussed how more general global symmetries of
multipartite systems constrain the entanglement between two
neighboring particles. We prove that  maximal entanglement
(measured in terms of the entanglement of formation) is always
attained for the ground state of a certain nearest neighbor
interaction Hamiltonian having the considered symmetry with the
achievable amount of entanglement being a function of the ground
state energy. Systems of Gaussian states, i.e. quantum harmonic
oscillators, are investigated in more detail and the results are
compared to what is known about ordered qubit systems.
\end{abstract}


%


\section{Introduction}
Entanglement is the type of correlations which can be shared only
with a finite number of parties---or to express it with the words
of C.H. Bennett: ''Entanglement is monogamous`` (cf.
\cite{Bennettquote}). This is maybe one of the main
characteristics of entanglement and it clearly distinguishes
entanglement from classical correlations. The present paper is
devoted to investigate this ''monogamy property`` of entanglement
in symmetric multipartite systems with the particular focus on the
relation to frustration, the existence of local hidden variable
models and to ground states of Hamiltonians having the considered
symmetry. The article is based on a talk given at the QIT-EQIS
workshop in Kyoto 2003 and it essentially reviews and extends
results from \cite{WVC03} and \cite{F82,W89,TDS03}.

In order to understand a characteristic feature of entanglement it
is useful to study the counterpart in classical correlations
first---this will clarify the difference between the quantum and
the classical world. To this end Sec.\ref{SecBell} will as a
starting point discuss frustration in classical systems and their
relation to Bell inequalities. In Sec.\ref{SecExtensions} we will
then utilize these observations in order to prove in a simple way
that there exists a local hidden variable model for quantum states
having symmetric extensions. Sec.\ref{SecEnergy} considers more
general symmetries and  investigates the relation between the
problem of maximizing the entanglement between nearest neighbors
under a global symmetry constraint (e.g. translational symmetry)
and the task of calculating ground state energies of certain
nearest neighbor interaction Hamiltonians. Finally,
Sec.\ref{SecGauss} will apply the ideas of the preceding section
to Gaussian systems with symmetries characterized by symmetric
graphs.

\section{Bell inequalities and frustration in classical
systems}\label{SecBell}

Constraints on the possible range of correlations in the form of
inequalities have been investigated for many years, even before
physicists developed an interest in that subject due to the work
of Bell \cite{B64} (see the monograph by Fr\'echet \cite{F40}).
The relation between Bell inequalities and frustration in
classical systems was then pointed out and studied in the early
eighties in particular by Fine \cite{F82}.

In spite of the simplicity of this connection there are, however,
surprisingly many publications in the field of quantum information
theory in which this knowledge has apparently disappeared. The
following section will recall the relations between joint
distributions, local hidden variable models, frustration and Bell
inequalities.

\subsection{Frustration in classical systems is due to loops}

Let us consider a set of observables $X_1,\ldots,X_n$ described
within classical probability theory and let us denote the
probability that observable $X_i$ leads to the outcome $x$ by
$P_{X_i}(x)$. Obviously, there exists always a \emph{joint
probability distribution} $P_{X_1,\ldots,X_n}=\prod_i P_{X_i}$
which returns all the single distributions as marginals. However,
if we fix in addition the pair distributions $P_{X_i, X_j}$ for a
certain subset of pairs $(i,j)$ in a non-trivial way, a joint
distribution with these marginals in general no longer exists. A
necessary condition for the existence of $P_{X_1,\ldots,X_n}$ is
of course the \emph{compatibility} of overlapping distributions in
the sense that \eq{\label{EqCompatibility}\sum_{x_2}
P_{X_1,X_2}(x_1,x_2) = \sum_{x_3} P_{X_1,X_3}(x_1,x_3) =
P_{X_1}(x_1).} However, it is in general not sufficient, and the
standard counter-example is a set of three observables with three
pair distributions corresponding to total anti-correlations.
\begin{figure}[t]\begin{center}
\epsfig{file=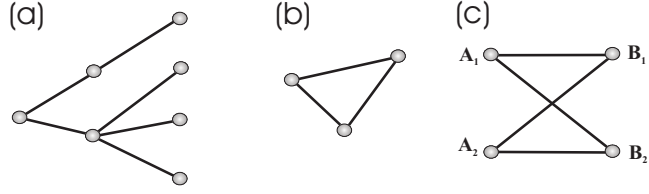,width=8.5cm} \caption{Frustration in
classical systems is due to loops. Let the vertices and edges of a
graph be assigned to observables and given pair distributions
(correlations) respectivley. Then (a) for tree graphs there exists
always a joint distribution returning all the given distributions
as marginals, (b) loops may cause frustration and (c) a necessary
and sufficient condition for the existence of a joint distribution
in a two-colorable graph is given by Bell's inequalities.}
\label{FigGraphs1}
\end{center}\end{figure}

It is very useful to depict the problem graphically and to assign
a vertex to every observable and an edge to each of the fixed pair
distributions. In this picture, compatibility of the pair
distributions is sufficient for the existence of a joint
distribution $P_{X_1,\ldots,X_n}$ if the graph does not contain
any loops:

\begin{proposition}\label{PropLoops}
Consider a graph, such that every vertex $k$ corresponds to one of
$n$ observables $X_k$ and every edge $(k,l)$ corresponds to a
given pair distribution $P_{X_k,X_l}$. If all overlapping
distributions are compatible in the sense of
Eq.(\ref{EqCompatibility}) and the graph does not contain any
loop, then there exists a joint probability distribution
$P_{X_1,\ldots,X_n}$ which returns all the given distributions as
marginals.
\end{proposition}
\proof{ The proof can easily be formulated in terms of an
induction. Assume there exists already a joint distribution
$P_{X_1,\ldots,X_i}$ which is compatible with all fixed pair
distributions $P_{X_k,X_l}$ where $k,l\leq i$ and has the same
marginal $P_{X_i}$ as all given pair distributions $P_{X_i,X_j}$
where $j>i$. Then we can construct a new distribution
\eq{P_{X_1,\ldots,X_{i+1}}=P_{X_1,\ldots,X_i}\; P_{X_i,X_{i+1}}\;
P^{-1}_{X_i}}which has all the required properties for a set of
$i+1$ vertices. Since each step in the induction adds one edge and
one vertex to the graph corresponding to the previous step, the
construction only works for tree graphs, i.e., graphs without
loops---otherwise one would at some point have to add an edge
without adding a new vertex.

The final joint distribution is then given by
\eq{P_{X_1,\ldots,X_n} = \left[\prod_{(k,l)}
P_{X_k,X_l}\right]\cdot\left[\prod_j P^{(1-e_j)}_{X_j}\right],}
where $e_j$ is the number of edges at the $j$'th vertex and the
products have to be taken over all edges $(k,l)$ and vertices $j$
respectively\footnote{Note that the extension is not unique.
Consider for instance a three-vertex graph with edges $(1,2)$,
$(1,3)$ and joint probability $P_{X_1, X_2, X_3}$. In this case we
could as well take $\tilde{P}_{X_1, X_2, X_3}=P_{X_1, X_2, X_3}+Q$
where $Q$ is any function fulfilling
$\sum_{x_2}Q(x_1,x_2,x_3)=\sum_{x_3}Q(x_1,x_2,x_3)=0$ and $P_{X_1,
X_2, X_3}+Q\geq 0$.}. }

 Prop. \ref{PropLoops} shows that
classically frustration is due to loops. This is in contrast to
quantum mechanical systems, which can be frustrated even without
loops.

\subsection{Joint distributions and local hidden variable models}

If the graph characterizing the fixed correlations between pairs
of observables is \emph{two-colorable}\footnote{A graph is called
\emph{two-colorable}, \emph{bicolorable} or \emph{bipartite} if we
can divide the set of vertices into two disjoint sets such that no
two vertices within the same set are connected by an edge. This is
equivalent to saying that all the cycles are of even length.} like
in Fig.\ref{FigGraphs1}(c) then a necessary and sufficient
condition for the existence of a joint distribution is given by
the complete set of Bell inequalities, i.e., by the existence of a
local hidden variable model. In this sense the correlations of a
quantum state which violates a Bell inequality lead to
unresolvable frustration when described within classical
probability theory.

Let us briefly recall the above mentioned result. We say that a
set of measured correlations $\big\{P_{A_iB_j}\big\}$ between
observables $A_1,\ldots, A_{m_A}$ and $B_1,\ldots, B_{m_B}$ admit
a description within a local hidden variable model \cite{F82,
WW01, M93}, if we can write
\eq{\label{lhiv2prob}P_{A_iB_j}(a,b)=\int_{\Lambda} M(d\lambda)
\chi_{A_i}(a,\lambda) \chi_{B_j}(b,\lambda).} Here
$\lambda\in\Lambda$ is the \emph{hidden variable} and the source
of the correlation experiment is characterized by the
probabilities with which the different $\lambda$ occur, i.e., by a
probability measure $M$ on $\Lambda$. The \emph{response function}
$\chi_{A_i}(a,\lambda)$ gives the probability that measuring the
system in state $\lambda$ with observable $A_i$ leads to the
outcome $a$. The locality assumption in Eq.(\ref{lhiv2prob}) is
expressed in the fact that the response functions factorize, such
that $\chi_{A_i}(a,\lambda)$ does not depend on $B_j$ and
$\chi_{B_j}(b,\lambda)$ is independent of $A_i$.

The local hidden variable description of a set of correlations is
equivalent to the existence of a joint probability distribution
\cite{F82}:
\begin{proposition}
\label{PropBellJoint} There exists a joint probability
distribution for all given pair distributions
$\big\{P_{A_iB_j}\big\}$ if and only if the correlations admit a
description within a local hidden variable model.
\end{proposition}
\proof{ Let $A=(A_1,\ldots,A_{m_A})$, $a=(a_1,\ldots,a_{m_A})$ be
vectors of observables and their respective outcomes, and
similarly for $B,\ b$. If $P_{A, B}$ is the joint distribution for
all pair distributions $P_{A_i, B_j}$, then
\begin{equation}\label{jointmodel}
P_{A_i,B_j}(\alpha,\beta)=\sum_{a, b} P_{A, B}(a,b)
\;\delta_{\alpha, a_i} \;\delta_{\beta, b_j}
\end{equation}is an admissible (deterministic) local hidden
variable model with $P_{A,B}$ playing the role of the measure $M$
and the two delta functions corresponding to the characteristic
functions in Eq.(\ref{lhiv2prob}).

Conversely, if $\big(\Lambda$, $M$, $\{\chi_{A_i}\}$,
$\{\chi_{B_j}\}\big)$ are probability space, measure and response
functions for a local hidden variable model, then
\begin{equation}\label{modeljoint}
P_{A, B}(a,b) \equiv \int\! M(d\lambda) \prod_{i=1}^{m_A}
\chi_{A_i}(a_i,\lambda) \prod_{j=1}^{m_B} \chi_{B_j}(b_j,\lambda)
\end{equation} is a joint distribution, which returns all pair
distributions $P_{A_i,B_j}$ as marginals. }

Note that Prop.\ref{PropBellJoint} naturally generalizes to
$n$-partite systems, where correlations of the form
$P_{A_iB_j\ldots N_k}$ instead of pair distributions are given.

\section{Quantum states with symmetric
extensions}\label{SecExtensions}

As already mentioned frustration in quantum systems are possible
even without loops in the configuration. The simplest example is a
system of three qubits distributed to Alice, Bob and Charlie,
where Alice wants to be maximally entangled with both her
colleagues. Clearly, this is not possible and one way to see this
is to realize that in such a scenario Alice could teleport an
unknown qubit perfectly to both, Bob and Charlie, which
contradicts the no-cloning theorem\cite{WZ82}. In fact, if Alice
is maximally entangled with Bob she cannot share \emph{any}
correlations with Charlie, neither
 entanglement nor classical correlations and it has recently been
 proven that there is even a quantitative trade-off between the amount of
 entanglement Alice shares with Bob and the maximal amount of classical
 correlations shared with Charlie \cite{KW03}.

\begin{figure}[t]\begin{center}
\epsfig{file=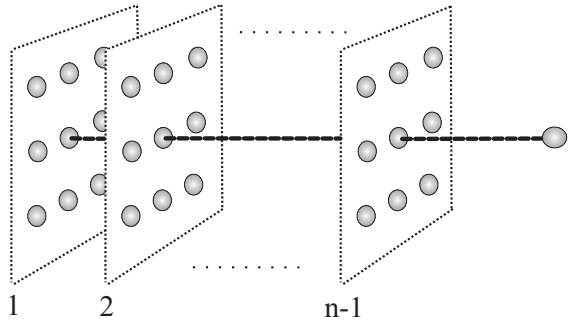,width=7.5cm} \caption{Consider a
multipartite quantum systems composed out of a single particle (on
the right) and $(n-1)$ ``layers'' of $m$ particles each. Assume
complete permutation symmetry within each layer and let $\rho_n$
be the $n$-partite reduced state containing one particle of each
layer plus the remaining single particle. Then $\rho_n$ admits a
local hidden variable description for all correlations of an
arbitrary number of observables on the $n$'th site and $m$
observables on each of the first $(n-1)$ sites.} \label{Figlayers}
\end{center}\end{figure}

 We will in the following restrict to symmetric situations, where
 we have for instance  Alice being surrounded by several Bobs
 and Charlies
 in a symmetric manner. Using the observations of the previous
 section, we can easily recover  results of \cite{W89} and \cite{TDS03}:

\begin{proposition}\label{PropBellExt}
 Let $\rho$ be a density matrix acting on a
 Hilbert space ${\cal H}_A\otimes{\cal H}_B^{\otimes m}$ such that
 all bipartite reduced states of the form $\rho_{AB_j}$ are
 equal. Then the state $\rho_{AB}\equiv\rho_{AB_j}$ admits a local hidden variable model with respect to all
 correlations between $m$ observables on site $B$ and an
 arbitrary number of observables measured on site $A$.
 \end{proposition}
\proof{ Let $\{F_i(a_i)\}$, $\{G_j(b_j)\}$ be the POVM operators
corresponding to the measurement devices $A_i$ and $B_j$ with
respective measurement outcomes $a_i$ and $b_j$. Then
\begin{equation}\label{extjoint}
  P_{A_i,B_1,\ldots,B_{m}}(a_i,b_1,\ldots,b_{m})=\tr{\rho F_i(a_i)\otimes\bigotimes_{j=1}^{m} G_j(b_j)}
\end{equation}
is an admissible joint probability distribution compatible with
all pair distributions $P_{A_i,B_j}(a_i,b_j)=\tr{\rho_{AB}
F_i(a_i)\otimes G_j(b_j)}$. By applying Prop.\ref{PropLoops} and
Prop.\ref{PropBellJoint} we have then that there exists a local
hidden variable model for all correlations of an arbitrary number
of observables acting on $A$ and $m$ measurement devices for $B$.
}

Prop. \ref{PropBellExt} shows that if the number $m$ of equal
neighbors in a quantum system is increased, then the bipartite
correlations between the central particle and one of the neighbors
become more and more classical. In fact, is was conjectured by
Schumacher and proven by Werner (cf.\cite{W89, DPS03}) that if $m$
tends to infinity, then $\rho_{AB}$ cannot contain any
entanglement---this is the only possibility of having a starlike
extension with arbitrary many parties.

The following result generalizes Prop.\ref{PropBellExt} to cases
in which local hidden variable models for more than two parties
can be constructed:

\begin{proposition}\label{PropBellExtMulti}
Let $\rho$ acting on ${\cal H}^{\otimes N}$ be a density operator
characterizing a state of $N=(n-1)m+1$ particles with permutation
symmetry within each of $(n-1)$ disjoint sets of $m$ particles
(see Fig.\ref{Figlayers}). Take the $n$-partite reduced state
$\rho_n$ which contains one particle of each set plus the
remaining particle. Then $\rho_n$ admits a local hidden variable
description for all correlations of an arbitrary number of
observables on the $n$'th ``single particle'' site and $m$
observables on each of the remaining $(n-1)$ sites.
\end{proposition}
\proof{ Let $\big\{F_j^{(i)}(a_j^{(i)})\big\}$ be the POVM
corresponding to the $j$'th observable on site $i$ with respective
outcome $a_j^{(i)}$. Then \eq{\label{manytensors}\tr{\rho
\bigotimes_{k=1}^m\bigotimes_{i=1}^{n-1}
F_k^{(i)}(a_k^{(i)})\otimes F_j^{(n)}(a_j^{(n)})}} is an
admissible joint distribution for one observable ($j$) on the
$n$'th site and $m$ observables on the remaining $(n-1)$ sites.
Having the tensor factors properly ordered, this distribution is
 compatible with all correlations measured on
$\rho_n$. This is seen by noting that taking the sum over
$a_k^{(i)}$ in Eq.(\ref{manytensors}) corresponds to tracing out
the $k$'th particle in the $i$'th set (``layer'' in
Fig.\ref{Figlayers}). Due to the assumed permutation symmetry of
$\rho$ it does, however, not matter which $(m-1)$ particles in
each layer are traced out---we will always end up with the
$n$-partite reduced state $\rho_n$.

By Prop.\ref{PropLoops} the probability distribution in
Eq.(\ref{manytensors}) can be extended to a joint distribution
including an arbitrary number of observables on the $n$'th site
and by the multipartite generalization of Prop.\ref{PropBellJoint}
there exists a local hidden variable model for all the considered
correlations.\footnote{Note that the conditions of
Prop.\ref{PropBellExtMulti} can be weakened in two directions:
First, permutation symmetry within each layer is not required---it
is sufficient that all possible $n$-partite reduced states are
equal to $\rho_n$. Second, as mentioned in Ref.\cite{TDS03},
positivity of $\rho$ is only required on product operators.
Moreover, the numbers of particles per layer may be different.} }

\section{Maximizing entanglement and minimizing
energy}\label{SecEnergy}

Prop. \ref{PropBellExt} and Prop. \ref{PropBellExtMulti} show that
an increasing symmetry of a quantum state (in the sense of an
increasing number of equal neighbors) constrains the entanglement
 in such a way that violating Bell
inequalities of a certain type is no longer possible. In the
following we will consider multipartite quantum systems which have
 more general symmetries with respect to the ordering of the
particles and investigate quantitatively how this global symmetry
constraints the entanglement between two neighboring particles.

The considered symmetry group will always be a subgroup $G$ of the
 group ${\cal S}_n$ of all permutations of $n$ parties. The
 global density operator $\rho$ then commutes with all group
 elements\footnote{The group $G$ is represented by unitary operators $\{U_g\}$ which permute the tensor factors of the total Hilbert space ${\cal H}^{\otimes n}$.} and it is in particular  invariant under the group
 averaging
 \eq{\rho\mapsto {\cal T}(\rho):=\frac1{|G|}\sum_{g\in G} U_g \rho
 U_g^\dag\;.}
 It is in many cases again advantageous to depict the problem
 graphically and to assume that the group $G$ is the symmetry
 group of a graph whose vertices correspond to particles, i.e., tensor factors of the total Hilbert space.
  For simplicity we will only consider \emph{edge transitive
 graphs} here\footnote{That is, every edge can be
 mapped onto every other edge by an element of the symmetry group.}. This provides a natural notion of ``neighboring''
 particles, namely those corresponding to adjacent vertices in the
 graph, and we will not have to specify which neighbors we are considering since they are all equal.
 Prominent examples of edge transitive graphs are stars, rings,
 cubic, hexagonal and trigonal lattices, permutational invariant clusters and the platonic solids.

Typically, states with these symmetries appear as ground states
(or equilibrium states) of particles on a lattice with equal
nearest neighbor interaction along the edges of the considered
graph\footnote{if there is no symmetry breaking}. We will show
that the state which has the largest nearest neighbor entanglement
under such a symmetry constraint is always the ground state of a
certain nearest neighbor interaction Hamiltonian and that,
moreover, the maximal achievable amount of entanglement is a
function of the ground state energy.

This statement is rather obvious if we quantify the entanglement
using a linear functional like the the overlap
$f=\langle\Phi_-|\rho_{AB}|\Phi_-\rangle$ with the singlet state
$\Phi_-$ of two qubits.
 In fact, in this case the maximum is always achieved for
the ground state of the antiferromagnetic Heisenberg Hamiltonian,
since we can write
\begin{equation}\label{EqHeisenberg}
|\Phi_-\rangle\langle\Phi_-| = \frac14 \Big[{\bf 1} -
\sigma_x\otimes\sigma_x-\sigma_y\otimes\sigma_y-\sigma_z\otimes\sigma_z\Big].
\end{equation}
In this way we can for instance relate the ground state energy
density of the infinite antiferromagnetic Heisenberg
spin-$\frac12$ chain to the maximal achievable singlet fraction of
a state with infinite translation symmetry, which is then given by
$f_{max}=\ln 2$.\footnote{Bounds on the maximally entangled
fraction under symmetry constraints will be studied in greater
detail in \cite{EWW03}.}

However, an analogous statement is still true if we measure the
entanglement in terms of a highly nonlinear functional, the
\emph{entanglement of formation}\cite{BDSW96}. In contrast to the
singlet fraction the entanglement of formation is a proper
\emph{entanglement measure} and it is closely related to the
amount of pure state entanglement needed to prepare a state by
means of local operations and classical communication. It is
defined as \eq{E_F(\rho_{AB})=\inf\left\{\sum_i p_i
E(\Psi_i)\Big|\sum_i
p_i|\Psi_i\rangle\langle\Psi_i|=\rho\right\}\;,\label{EqEofDef}}
where $E(\Psi)=S\big({\rm
tr}_A\big[|\Psi\rangle\langle\Psi|\big]\big)$ is the pure state
entanglement of $\Psi$ given by the von Neumann entropy $S$  of
the reduced state. Due to the fact that Eq.(\ref{EqEofDef}) is a
so-called \emph{convex hull} construction, we can prove the
following:
\begin{proposition}\label{PropHvsEF}
Consider the set ${\cal D}_G$ of multipartite quantum states with
(permutation) symmetry group $G$ of an edge transitive graph. The
maximal entanglement $E_F(\rho_{AB})$ between two neighboring
particles of a state $\rho\in{\cal D}_G$ is attained for the
ground state projector of a nearest neighbor interaction
Hamiltonian with interactions along the edges of the considered
graph. The maximal achievable entanglement is then a function of
the ground state energy.
\end{proposition}
\proof{ First note that by the concavity of the entropy we can
take the infimum in Eq.(\ref{EqEofDef}) over all decompositions of
$\rho_{AB}$ into mixed states $\{\sigma_i\}$ as well. Moreover,
$E_F$ is by construction the convex hull of the functional
$x\mapsto S\big({\rm tr}_A[x]\big)$, which can equivalently be
expressed as the supremum over all affine functions, which lie
below it (cf.\cite{AB03}). That is
\begin{eqnarray}
E_F(\rho_{AB}) &=& \inf_{\{p_i,\sigma_i\}}\left\{\sum_i p_i
S\big({\rm
tr}_A[\sigma_i]\big)\Big|\sum_i p_i\sigma_i=\rho\right\}\\
&=& \sup_h\left\{\tr{\rho_{AB}h}\Big|\forall\sigma: \tr{\sigma
h}\leq S\big({\rm tr}_A[\sigma]\big)\right\}\\
&=& -\inf_h \left\{\tr{\rho_{AB}h}\Big|\forall\sigma: \tr{\sigma
h}+ S\big({\rm tr}_A[\sigma]\big)\geq 0\right\}\label{EqDefhs}\\
&=& -\inf_s \tr{\rho_{AB}h(s)},
\end{eqnarray}
where $h(s)$ is a fictive parametrization of the set of
``interactions'' $h$ defined in Eq.(\ref{EqDefhs}).

By assumption $\rho_{AB}$ is a bipartite reduced state of a global
state $\rho\in{\cal D}_G$ which is invariant under averaging over
the group $G$. We can therefore write
\begin{eqnarray}
\tr{\rho_{AB}h(s)} &=& \tr{\rho\big(h(s)\otimes{\bf 1}\big)}=
\tr{{\cal T}\big(\rho\big)\big(h(s)\otimes{\bf 1}\big)}\\
&=& \tr{\rho{\cal T}\big(h(s)\otimes{\bf 1}\big)}=:\tr{\rho H(s)},
\end{eqnarray}
where $H(s)$ is a ``Hamiltonian'' with equal nearest neighbor
interactions $h(s)$ along the edges of the considered graph. Due
to the symmetry of $H(s)$ we can now drop the constraint
$\rho\in{\cal D}_G$ in calculating the maximal achievable
$E_F(\rho_{AB})$:
\begin{eqnarray}
\sup_{\rho\in{\cal D}_G} E_F(\rho_{AB}) &=& -\inf_s \inf_{\tau}
\tr{\tau H(s)} = -\inf_s e_0(s).
\end{eqnarray}
Here, $e_0$ is the ground state energy of the Hamiltonian $H(s)$
and the normalized projector onto the ground state space of the
extremal Hamiltonian has then both, the required symmetry and the
maximal entanglement properties. }

There are two drawbacks concerning the application of
Prop.\ref{PropHvsEF}. First of all, the parametrization of $h(s)$
is in general not known explicitly. In fact, there are only very
few systems for which we are able to calculate $E_F(\rho_{AB})$.
Besides highly symmetric one-parameter families of states
\cite{VW01, TV00} this is at present only feasible for systems of
two qubits \cite{W98} and symmetric two-mode Gaussian states
\cite{GWKWC03}.

The second problem is the calculation of the ground state energy
$e_0(s)$ for large systems. Though some particular two-qubit
interactions lead to exactly solvable models in one dimension, the
set of these models is apparently not large enough in order to
answer the question about the maximal possible $E_F(\rho_{AB})$
for an infinite qubit chain in a straight forward manner. Using
results from Ref.\cite{VDM02} we can write the \emph{concurrence}
of two qubits (which is in turn a monotone function of $E_F$
\cite{W98}) as
\eq{\label{Concurrence}c\big(\rho_{AB}\big)=\max\left\{0,-\inf_{\det{X}=1}
\tr{\rho_{AB} (X\otimes X^\dag {\bf F})}\right\},} where ${\bf F}$
is the \emph{flip operator}\footnote{The flip operator
interchanges the two tensor factor: ${\bf
F}|\phi\rangle\otimes|\psi\rangle=|\psi\rangle\otimes|\phi\rangle$}.
For a chain of qubits this leads to a somehow deformed XXZ +
Z-field model which is, unfortunately, not exactly solvable.

However, for Hamiltonians which are quadratic in bosonic
operators, i.e., interactions with Gaussian ground states, both
problems---calculating $E_F$ (that is parameterizing $h(s)$) and
calculating $e_0(s)$---are feasible.

\begin{figure}[t]\begin{center}
\epsfig{file=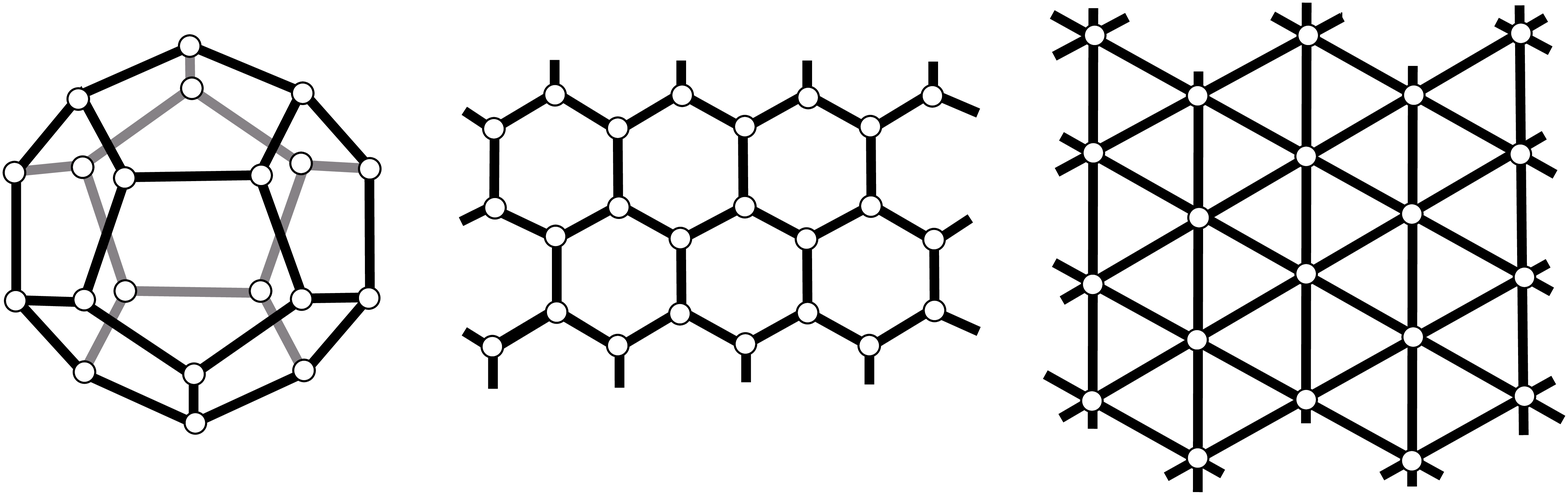,width=9cm} \caption{Examples of
symmetric graphs are the five platonic solids (e.g. the
dodecahedron) and the hexagonal and trigonal lattice. If every
vertex corresponds to a single mode of a Gaussian state then the
entanglement $E_F$ between nearest neighbors is maximized for the
ground state of the nearest neighbor Hamiltonian given in
Eq.(\ref{EqHamOpt}). Although the entanglement is finite in each
case, the respective ground states are infinitely squeezed.}
\label{FigSymGraphs}
\end{center}\end{figure}

\section{Gaussian states on symmetric graphs}\label{SecGauss}

Let us consider a bosonic system of $n$ modes described by a set
of canonical operators $(Q_1,\ldots,Q_n,P_1,\ldots,P_n)=:R$
obeying the canonical commutation relations $[Q_k,P_l]=i
\delta_{kl}{\bf 1}$. States with Gaussian Wigner
distribution---so-called \emph{Gaussian states}---are completely
characterized by their first and second moments with respect to
the canonical operators \cite{H82}. Physically they may describe
modes of the electromagnetic field, atomic ensembles interacting
with such fields, the motional state of a collection of ions in a
trap or the low energy (bosonic) excitations of many other
systems. All the information about correlations and entanglement
properties of these states is contained in the \emph{covariance
matrix} \eq{\label{EqCovMat}\Gamma_{kl}=\tr{\rho\Big\{R_k-\langle
R_k\rangle,R_l-\langle R_l\rangle\Big\}_+},}where $\langle
R_k\rangle =\tr{\rho R_k}$ and $\{\cdot,\cdot\}_+$ is the
anti-commutator.

It was proven in Ref.\cite{WVC03} that for the case of Gaussian
states on symmetric graphs\footnote{\emph{Symmetric graphs} are
those which are edge transitive and vertex transitive. Examples
are rings,
 cubic, hexagonal and trigonal lattices, permutational invariant clusters and the platonic
 solids (see Fig.\ref{FigSymGraphs}).}, where every vertex corresponds to a single mode (one quantum harmonic
 oscillator), the Hamiltonian whose ground
 state maximizes $E_F(\rho_{AB})$ is of the form
 \begin{eqnarray}
 H &=& \sum_{(k,l)} (Q_k+Q_l)^2+(P_k-P_l)^2\label{EqHamOpt}\\
 &=& \sum_{(k,l)} (a_k^\dag a_k+\frac12)+\frac12(a_ka_l+a_k^\dag
 a_l^\dag)\\
 &=:& \sum_{ij} h_{ij} R_i R_j\,
 \end{eqnarray}
 where the first two sums run over all edges $(k,l)$ of the graph and the \emph{Hamiltonian
 matrix} $h\equiv h_Q\oplus h_P$ is block diagonal with the blocks
 $h_Q$ and $h_P$ being easily determined from the adjacency matrix
 of the graph. The ground state energy\footnote{The relation between $e_0$ and the maximal achievable entanglement $E_F$ is in this case different than suggested in the proof of Prop.\ref{PropHvsEF}. However, $E_F$ is still given by a monotone decreasing function of $e_0$.} $e_0$ of the Hamiltonian
 and the covariance matrix of the respective ground state $\Gamma_0$
 are then given by
 \eq{e_0=||\sqrt{h_Qh_P}||_1\;,\qquad
 \Gamma_0=\sqrt{h_Ph_Q^{-1}}\oplus\sqrt{h_Qh_P^{-1}}.}
 Although the entanglement is finite in all non-trivial
 cases\footnote{If we apply the interaction in Eq.(\ref{EqHamOpt})
 to only two particles, then the ground state will be the
 infinitely entangled original EPR state \cite{EPR35}.} all the ground states
 are \emph{infinitely squeezed}, which is mathematically expressed
 in the fact that $\Gamma_0$ has zero eigenvalues.

\begin{figure}[t]\begin{center}\psfrag{e}{$E_F(\rho_{AB})$}\psfrag{n}{N}\psfrag{5}{\small 5}\psfrag{10}{\small 10}\psfrag{15}{\small 15}
\psfrag{20}{\small 20}\psfrag{0.1}{\small\hspace{4pt}
0.1}\psfrag{0.2}{\small\hspace{4pt}
0.2}\psfrag{0.3}{\small\hspace{4pt}
0.3}\psfrag{0.4}{\small\hspace{4pt}
0.4}\psfrag{0.5}{\small\hspace{4pt} 0.5}
\epsfig{file=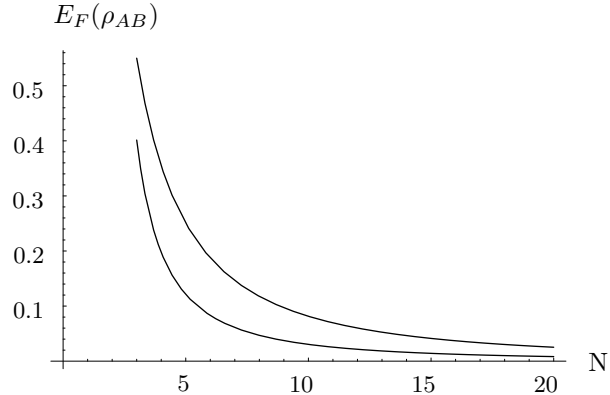,width=9cm} \caption{The maximal
achievable entanglement $E_F$ (ebits) between two particles in an
$N$-partite permutational symmetric cluster. The lower curve
corresponds to $N$-mode Gaussian states, and the upper curve shows
the result for an $N$-qubit system. Both vanish asymptotically as
$\sim\frac1{N^2}\log_2{N}$.} \label{FigCluster}
\end{center}\end{figure}

\begin{figure}[t]
\psfrag{E}{\hspace{-5pt}$E_F(\rho_{AB})$}\psfrag{0.3}{\small
0.3}\psfrag{0.4}{\small 0.4}\psfrag{0.5}{\small
0.5}\psfrag{0.6}{\small 0.6}\psfrag{3}{\small 3}\psfrag{6}{\small
6}\psfrag{9}{\small 9}\psfrag{12}{\small 12}\psfrag{15}{\small
15}\psfrag{18}{\small 18}
\psfrag{e}{$E_F(\rho_{AB})$}\psfrag{n}{N}
\epsfig{file=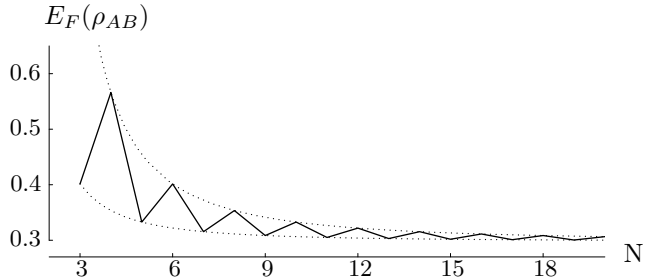,width=10.5cm} \caption{Maximal nearest
neighbor entanglement (ebits) in a ring of $N$ harmonic
oscillators (the dotted curves represent the envelopes). In the
limit $N\rightarrow\infty$ this approaches $0.30$ ebits, which is
comparable to the $0.29$ ebits Wootters found as a lower bound for
the case of an infinite qubit chain.\label{FigEntChain}}
\label{FigChain}
\end{figure}

 Figs. \ref{FigChain} and \ref{FigCluster} show the (analytic) results for a translational
 invariant ring and a permutational invariant cluster of harmonic
 oscillators and compare them to what is known from the case of
 qubit systems\footnote{The analytic result for the permutational invariant qubit cluster was derived in Ref.{\cite{KBI00}}. The case of qubit rings was investigated in Ref.\cite{CW01}, where a lower bound on the maximal nearest neighbor entanglement was derived.}. Surprisingly, the maximal value of $E_F$ is of the
 same order of magnitude for both systems and as it is  expected for a
 ring of qubits, the ring of harmonic oscillators shows an
 odd-even oscillation with respect to the number of particles.

 This together with other results from Ref.\cite{WVC03} indicates three
 different tendencies for the maximal $E_F$:

 \begin{enumerate}
    \item It decreases with the number of adjacent vertices.
    \item It decreases with the total number of  vertices.
    \item It is suppressed in loops with an odd number of
    vertices, which give thus rise to  additional frustration.
\end{enumerate}

\section*{Acknowledgements}
MMW acknowledges interesting discussions with R.F. Werner, K.M.R.
Audenaert and M.B. Ruskai during the EQIS conference in Kyoto.
This work was supported in part by the E.C. (projects RESQ and
QUPRODIS) and the Kompetenznetzwerk
``Quanteninformationsverarbeitung'' der Bayerischen
Staatsregierung.


\appendix


\end{document}